\documentclass[twocolumn,preprintnumbers,amsmath,amssymb,secnumarabic,amssymb, nobibnotes, aps, pra]{revtex4}
\usepackage{pxfonts}
\usepackage{amsmath}
\usepackage{bm}
\usepackage{epsfig}
\usepackage{mathrsfs}
\def\be{\begin{equation}}
\def\ee{\end{equation}}
\def\bea{\begin{eqnarray}}
\def\eea{\end{eqnarray}}

\begin{document}
\title{Controlling laser spectra in a phaseonium photonic crystal using maser}
\author{Pankaj K. Jha$^{1,}$\footnote{Email: pkjha@physics.tamu.edu}}
\author{C. Y. Lee$^{2}$}
\author{C. H. Raymond Ooi$^{2,}$\footnote{Email: rooi@um.edu.my}}
\affiliation{$^{1}$Texas A\&M University, College Station, Texas 77843, USA\\
$^{2}$University of Malaya, 50603 Kuala Lumpur, Malaysia}
\date{\today}
\begin{abstract}
We study the control of quantum resonances in photonic crystals with electromagnetically induced transparency driven by microwave field. In addition to the control laser, the intensity and phase of the maser can alter the transmission and reflection spectra in interesting ways, producing hyperfine resonances through the combined effects of multiple scattering in the superstructure. 
\end{abstract}
\maketitle
\section{Introduction}
Optical properties of matter can be substantially modified and controlled by coherent excitations. Such systems can exhibit quantum coherence and interference effects, e.g., enhanced nonlinear effects\cite{Tewari_PRL_86}, electromagnetically induced transparency (EIT)\cite{Harris90, Boller91,Field91}, giant Kerr nonlinearity\cite{Schmidt99,Harris99,Wang01},  lasing  without inversion\cite{LWI1,LWI2,LWI3}, efficient nonlinear frequency conversions \cite{Jain96,Jain93}, coherence Raman scattering enhancement via maximum coherence in atoms\cite{Scully02} and molecules\cite{Sari04}, enhanced lasing\cite{Jha12,Sete12}, coherent Raman umklappscattering\cite{CRU}, photodesorption\cite{JhaAPL12} to name a few. Recently quantum coherence effects has been applied to a new domain on plasmonics and shown to benefit nanophotonics\cite{Dorfman13,PKJha13}.

Additional coherences in the presence of  a microwave field, in addition to the optical fields have gained much attention in the past decade\cite{Wilson05,Kosa00}. For example, microwave field has been utilized to envision novel effects like electromagnetically induced transparency with amplification(EITA)\cite{Joo10} in superconducting qubits, sub-wavelength atom localization\cite{Sahrai05}, circuit QED\cite{Murali04}, simultaneous slow and fast light\cite{Luo10}, gains without inversion in quantum systems with broken parities\cite{Jia10}, quantum storage\cite{Novikova07,Novikova07b,Gorshkov07}, sub-Raman generation\cite{Jha13}. 

On the other hand superstructures composed of exotic materials, such as superconductors in photonic crystals, show interesting optical features~\cite{Ooi00,Ooi2010}. Similarly photonic crystal composed of dielectric and controllable quantum coherence medium can provide new optical properties. Recently we analyzed one such superstructure, i.e. a superlattice composed of alternating layers dielectric and quantum coherence phaseonium~\cite{Scully92} medium driven by a laser field~\cite{Ooi10}. Alternating layers with different optical properties can also be achieved by counter-propagating control laser~\cite {And05} and temperature tuning of superconducting layers \cite{super T}.

In this paper, we have extended the idea of controlling quantum resonances in photonic crystals with electromagnetically induced transparency by incorporating of a microwave coupling between the forbidden Raman transition $(|c\rangle \leftrightarrow |b\rangle )$ as shown in Fig. 1, thus extending the regime of microwave excitation to a new domain. The optical fields couple the dipole allowed transitions $(|a\rangle \leftrightarrow |b\rangle)$ and $(|a\rangle \leftrightarrow |c\rangle )$. Similar scheme in atomic $(^{87}\text{Rb})$ vapor  has already been experimentally demonstrated to exhibit control over EIT~\cite{Li09}. Here we show that the transmission and reflection profiles from single interface, double interface and superlattice structure show interesting features in the presence of the microwave field than its absence~\cite{Ooi00,Ooi2010}. In a closed loop both the amplitude and the phase of the microwave field can be applied to manipulate the dynamics of the system. The main results of the paper is shown in Figs. 3 and 4. where we have plotted the reflection, transmission and dispersion from the superlattice for different choices of phase and amplitude respectively of the microwave field.

It is worth mentioning here that the single atom formalism is applicable to dilute phaseonium particles such as doped transparent crystal and the model enables us to focus on the physics due to EIT without being drown in the many body formalism.  The interplay between microwave resonance and multiple reflections and transmissions in multilayered structures produces novel resonant features. Our system looks like, but not exactly a distributed feedback laser (DFL)\cite{Yariv}. The mechanism behind EIT and quantum coherence with maser is linear and therefore it cannot be considered as a laser. The gain (R, T$>$1) is acquired from an external laser which controls the transmission and reflection spectra. This kind of coherent control is not available in DFL. 

The paper is organized as follows, in section II we present our model and derive the equation of motion using the master equation approach. In section III we present our finding of the numerical simulations for reflection, transmission and dispersion profiles from single interface, double interface and superlattice in the presence of a microwave field coupled lower levels. We conclude our results in section IV where we have also discussed the application this approach.

\section{Model and Equation of Motion}
Let us consider a three-level medium in $\Lambda $ (Raman) configuration.
Here $|a\rangle \leftrightarrow |c\rangle $ is excited by a control field $\text{E}_{c}$, while we probe the transition $|a\rangle \leftrightarrow
|b\rangle $ with the field $\text{E}_{p}$. We have also coupled the
transition $|c\rangle \leftrightarrow |b\rangle $ by a micro-wave field $\text{E}_{m}$. Let us define the (real) control, probe and microwave fields
respectively as
\begin{equation}\label{eq1}
E_{c}(z,t)=(1/2)\mathcal{E}_{c}(z,t)\exp [i(k_{c}z-\nu _{c}t)]+\text{c.c},
\end{equation}
\begin{equation}\label{eq1a}
E_{p}(z,t)=(1/2)\mathcal{E}_{p}(z,t)\exp [i(k_{p}z-\nu _{p}t)]+\text{c.c},
\end{equation}
\begin{equation}\label{eq1}
E_{m}(z,t)=(1/2)\mathcal{E}_{m}(z)\exp [i(k_{m}z-\nu _{m}t)]+\text{c.c}.
\end{equation}
The atom-field interaction is given by
\begin{equation}  \label{eq2}
\begin{split}
\mathscr{H}=&\hbar (\omega _{a}|a\rangle \langle a|+\omega _{b}|b\rangle\langle b|+\omega _{c}|c\rangle \langle c|)-  \left[ \wp _{ab}E_{p}|a\rangle \langle b|\right.\\
&\left.+\wp _{ac}E_{c}|a\rangle \langle c|+\wp _{cb}E_{m}|c\rangle \langle b|+\text{H.c}\right].
\end{split}
\end{equation}
\begin{figure}[t]
\label{Fig1} \centerline{\includegraphics[height=13cm,width=0.45\textwidth,angle=0]{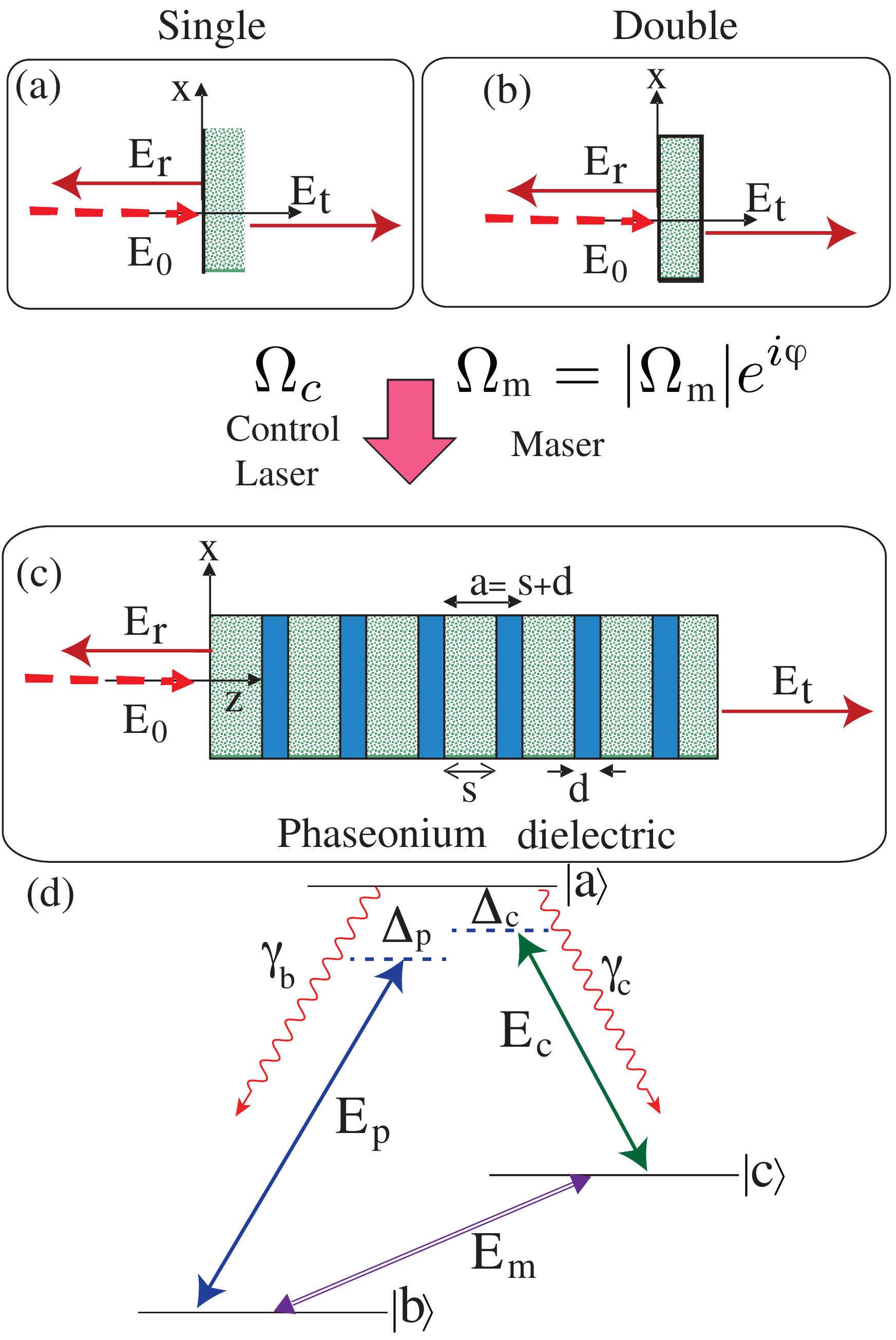}}
\caption{Plot of the schematic of the superlattice with alternating
layers composed of three-level scheme particles with EIT driven by control
laser and a maser. Here $E_{0}, E_{r}$ and $E_{t}$ are the incident, reflected and transmitted probe field $E_{p}$.}
\label{scheme}
\end{figure}
Equation of motion for the density matrix $\rho $ is given in
\begin{equation} \label{eq3}
\begin{split}
\frac{\partial \rho (z,t)}{\partial t} =-\frac{i}{\hbar }\left[ \mathscr{H},\rho \right] &+\frac{\gamma _{c}}{2}\left( [\sigma _{c},\rho \sigma_{c}^{\dagger }]+[\sigma _{c}\rho ,\sigma _{c}^{\dagger }]\right)\\ 
&+\frac{\gamma _{b}}{2}\left( [\sigma _{b},\rho \sigma _{b}^{\dagger}]+[\sigma _{b}\rho ,\sigma _{b}^{\dagger }]\right) ,
\end{split}
\end{equation}
where the atomic lowering ($\sigma _{q}$) and raising ($\sigma _{q}^{\dagger
}$) ($q=a,b,c$) operators are defined as $\sigma _{c}=\left| c\rangle \langle a\right| ,\sigma _{c}^{\dagger }=\left|
a\rangle \langle c\right| ;\,\sigma _{b}=\left| b\rangle \langle a\right|
,\sigma _{b}^{\dagger }=\left| a\rangle \langle b\right|$. To eliminate the fast oscillating terms we will use the following
transformation.
\begin{equation} \label{eq4a}
\rho _{ab}(t)=\varrho _{ab}\exp [i(k_{p}z-\nu _{p}t)], \\
\end{equation}
\begin{equation} \label{eq4b}
\rho _{ac}(t) =\varrho _{ac}\exp [i(k_{c}z-\nu _{c}t)], \\
\end{equation}
\begin{equation} \label{eq4c}
\rho _{cb}(t) =\varrho _{cb}\exp [i(\delta kz-\delta \nu t)].
\end{equation}
Here $\delta k=k_{p}-k_{c}$ and $\delta \nu =\nu _{p}-\nu _{c}$. The
evolution of the coherences $\varrho _{ij}$ takes the form
\begin{equation}\label{eq5a}
\begin{split}
\frac{\partial \varrho _{ab}}{\partial t} =&-\Gamma _{ab}\varrho_{ab}-i\Omega _{p}(\varrho _{aa}-\varrho _{bb})+i\Omega _{c}\varrho _{cb} \\
&-i\Omega _{m}\varrho _{ac}\exp [i(\Delta k_{m}z-\Delta \nu _{m}t)],\\
\end{split}
\end{equation}
\begin{equation}\label{eq5b}
\begin{split}
\frac{\partial \varrho _{ac}}{\partial t} =&-\Gamma _{ac}\varrho_{ac}-i\Omega _{c}(\varrho _{aa}-\varrho _{cc})+i\Omega _{p}\varrho_{cb}^{\ast}\\ 
&-i\Omega _{m}^{\ast }\varrho _{ab}\exp [-i(\Delta k_{m}z-\Delta \nu_{m}t)],\\
\end{split}
\end{equation}
\begin{equation}\label{eq5c}
\begin{split}
\frac{\partial \varrho _{cb}}{\partial t}=&-\Gamma _{cb}\varrho_{cb}+i\Omega _{c}^{\ast }\varrho _{ab}-i\Omega _{p}\varrho _{ac}^{\ast }\\
& -i\Omega _{m}(\varrho _{cc}-\varrho _{bb})\exp [i(\Delta k_{m}z-\Delta \nu_{m}t)],
\end{split}
\end{equation}
\noindent where we have defined the Rabi frequencies as $\Omega _{p}=\wp _{ab}\mathcal{E}_{p}/2\hbar,\Omega _{c}=\wp _{ac}\mathcal{E}_{c}/2\hbar,\Omega _{m}=\wp _{cb}\mathcal{E}_{m}/2\hbar$ and the detunings $\Delta k_{m}=k_{m}-\delta k,\,\Delta \nu _{m}=\nu _{m}-\delta \nu$. The decoherence $\Gamma_{ij}$ are given as
\begin{equation}
\Gamma_{ab}=(\gamma _{c}+\gamma _{b})/2+i(\omega _{ab}-\nu _{p})
\end{equation}
\begin{equation}
\Gamma_{ac}=(\gamma _{c}+\gamma _{b})/2+i(\omega _{ac}-\nu _{c}),
\end{equation}
\begin{equation}
\Gamma_{cb}=i(\omega _{cb}+\nu _{c}-\nu _{p}).
\end{equation}
Here we will consider the drive
field is on resonance with the transition $|a\rangle \leftrightarrow
|c\rangle $ i.e $\omega _{ac}=\nu _{c}$ while the probe and microwave field
are detuned by $\Delta _{p}$ and $\Delta _{m}$ respectively.
Analytical solution to Eqs. (\ref{eq5a}-\ref{eq5c}) can be found in
different regimes and various limits \cite{Thesis12}. Now assuming the drive field to be strong $|\Omega _{c}|\gg |\Omega
_{p}|,|\Omega _{m}|$ and $\Delta _{p}=\Delta _{m}$, let us solve Eqs.(\ref{eq5a}-\ref{eq5c}) in steady state regime ($\dot{\varrho}_{ab}=\dot{\varrho}_{ac}=\dot{\varrho}_{cb}=0$). We obtain,
\begin{equation}\label{eq6}
\begin{split}
\bar{\varrho}_{ab}=&i\frac{\Gamma _{cb}\Omega _{p}\left( |\Omega_{c}|^{2}n_{ac}/\Gamma _{ca}\Gamma _{cb}-n_{ab}\right) }{\Gamma _{ab}\Gamma_{cb}+|\Omega _{c}|^{2}}+\\
&\frac{\Omega _{c}\Omega _{m}\left( n_{cb}-\Gamma _{cb}n_{ac}/\Gamma_{ac}\right) \exp [i\Delta k_{m}z]}{\Gamma _{ab}\Gamma _{cb}+|\Omega_{c}|^{2}},
\end{split}
\end{equation}
where $n_{ij}=\varrho _{ii}-\varrho _{jj}$. If we assume $\varrho _{bb}\sim 1,\varrho
_{aa}=\varrho _{cc}\sim 0$, Eq.(\ref{eq6}) reduces to
\begin{equation}\label{eq7}
\bar{\varrho}_{ab}=i\frac{\Gamma _{cb}\Omega _{p}}{\Gamma _{ab}\Gamma
_{cb}+|\Omega _{c}|^{2}}-\frac{\Omega _{c}\Omega _{m}\exp [i\Delta k_{m}z]}{\Gamma _{ab}\Gamma _{cb}+|\Omega _{c}|^{2}}.
\end{equation}
Using the definition of complex susceptibility i.e $\chi =N|\wp _{ab}|^{2}\bar{\varrho}_{ab}/\hbar \epsilon _{0}\Omega _{p}$ and Eq.(\ref{eq7}), we obtain,
\begin{equation}
\begin{split}
\chi  =\frac{N|\wp _{ab}|^{2}}{\hbar \epsilon _{0}}\left[ i\frac{\Gamma
_{cb}\left( |\Omega _{c}|^{2}n_{ac}/\Gamma _{ca}\Gamma _{cb}-n_{ab}\right) }{\left( \Gamma _{ab}\Gamma _{cb}+|\Omega _{c}|^{2}\right) }\right. \\
 \left. +\frac{\Omega _{c}\Omega _{m}\left( n_{cb}-\Gamma
_{cb}n_{ac}/\Gamma _{ac}\right) \exp [i\Delta k_{m}z]}{\Omega _{p}\left(
\Gamma _{ab}\Gamma _{cb}+|\Omega _{c}|^{2}\right) }\right] .
\end{split}
\label{eq14}
\end{equation}
\noindent Eq.(\ref{eq14}) has two terms, the first term gives the well known
$\Lambda -$scheme EIT, and the second term is the contribution from the
microwave coupling of the transition $|c\rangle \leftrightarrow |b\rangle $.
The net effect is determined by the interference of these two terms. The
second term also determines the effect of relative phase of the optical and
the microwave field on the probe field transmission. Absolute phase effects in radio-frequency(RF) domain has also been demonstrated by Jha $et. al.$ \cite{Jha11a,Jha11b}. We use $k=\frac{\omega }{c}\sqrt{\epsilon _{q}(\omega )}$ with $\epsilon _{q}(\omega)=1+\chi (\omega )$ to obtain the dispersion of the probe field inside the
medium, $\omega $ versus Re$(k)$ and Im$(k)$, as shown in Fig.(2). Here we
\begin{figure}[t]
\label{Fig2} 
\centerline{\includegraphics[height=17cm,width=0.50\textwidth,angle=0]{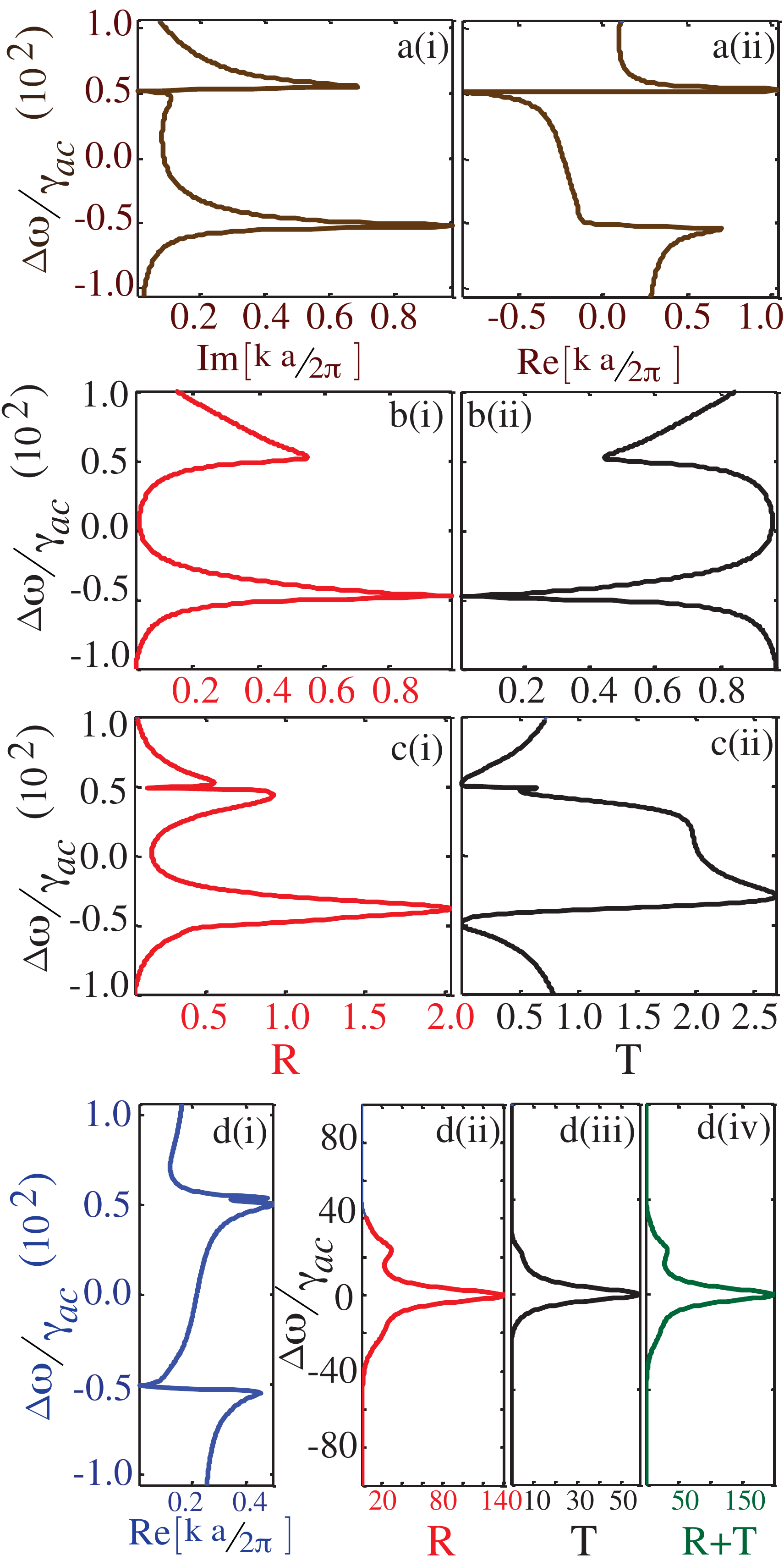}}
\caption{Plot of absorption and dispersion from double interface a(i)-(ii)
and dispersion from superlattice d(i). Reflection and transmission from single interface b(i)-(ii),
, double interface c(i)-(ii) and superlattice d(ii)-(iii) in the presence of the
maser field. For simulations we used $\Omega _{c}=50\protect\gamma _{ac}$, $\Omega _{p}=\protect\gamma _{ac}$, and $\Omega _{m}=0.02\protect\gamma _{ac}$, $N=10^{23}
$, $\protect\rho _{cc}=0.0296$, $\protect\rho _{aa}=0.0155$. Here $a=0.4\mu$m for the double interface (slab) and the superlattice and $d=s=0.5a$ for the superlattice with $n=8$.}
\end{figure}
\begin{figure*}[htb]
\centerline{\includegraphics[height=11.5cm,width=0.99\textwidth,angle=0]{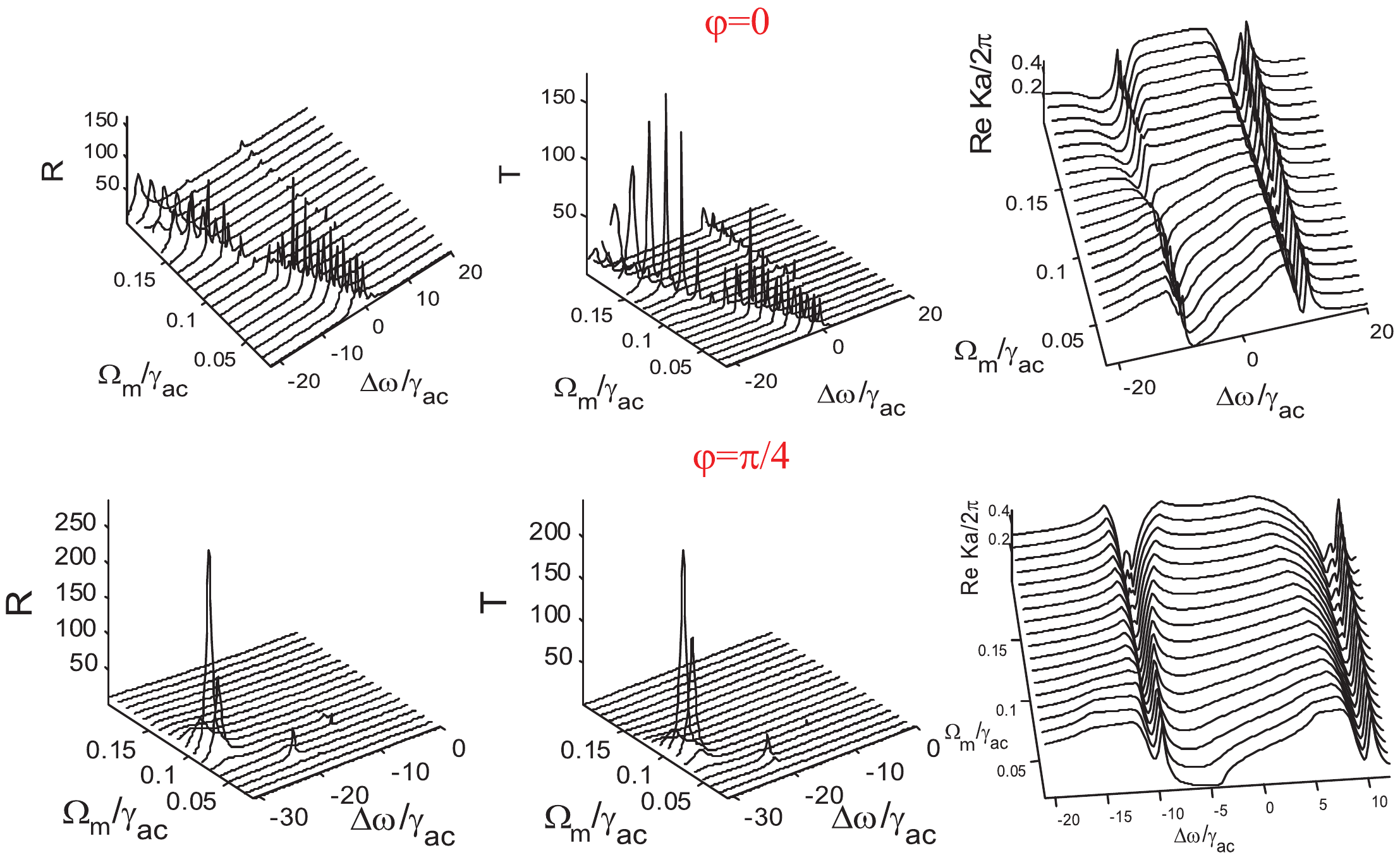}}
\caption{Plot of the reflection, transmission and dispersion versus the Rabi
frequency of the maser for a) $\protect\varphi =0$, b) $\protect\varphi =\protect\pi /4$, and c) $\protect\varphi =\protect\pi /2$. We use $\Omega_{c}=10\protect\gamma _{ac}$, $\Omega _{p}=0.1\protect\gamma _{ac}$, $\protect\rho _{cc}=0.0296$, $\protect\rho _{aa}=0.0155$ and other parameters are the same as in Fig. 2. Here we have defined $\varphi$ as the phase of the microwave field.}
\label{Fig3}
\end{figure*}
use $\omega $, a more common notation for spectral frequency, to replaces $\nu _{p}$ the probe frequency in Eq. (\ref{eq14}).
\section{Results}
To understand the effect of the microwave coupling on the transmission and
reflection we considered a single, double interface and superlattice. For
single interface between incident medium and quantum coherence medium $q$ we
use the Fresnel relations
\begin{equation}
\begin{split}
r_{iq}^{(p)}=\frac{\epsilon _{q}k_{iz}-\epsilon _{i}k_{qz}}{\epsilon
_{q}k_{iz}+\epsilon _{i}k_{qz}}, \quad
\end{split}
\end{equation}
\begin{equation}
\begin{split}
t_{iq}^{(p)}=\frac{2\sqrt{\epsilon _{q}\epsilon _{i}}k_{iz}}{\epsilon
_{q}k_{iz}+\epsilon _{i}k_{qz}}
\end{split}
\end{equation}
For double interface
\begin{equation}
\begin{split}
r_{io}^{^{\prime \prime }}=\frac{ r_{iq}+r_{qo}f^{2}}{1+r_{iq}r_{qo}f^{2}},
\end{split}
\end{equation}
\begin{equation}
\begin{split}
t_{io}^{^{\prime \prime }}=\frac{t_{iq}+t_{qo}f^{2}}{1+r_{iq}r_{qo}f^{2}}
\end{split}
\end{equation}
where $f=\exp(ik_{qz}d)$. The transmittance for single interface, $T=\left(
k_{iz}|t_{iq}^{2}|\right) /\left( k_{qz}\right) $, and for the double
interface, $T=|t_{io}^{^{\prime \prime }}|^{2}$.
\begin{figure*}
\centering
\includegraphics[height=11.5cm,width=0.99\textwidth,angle=0]{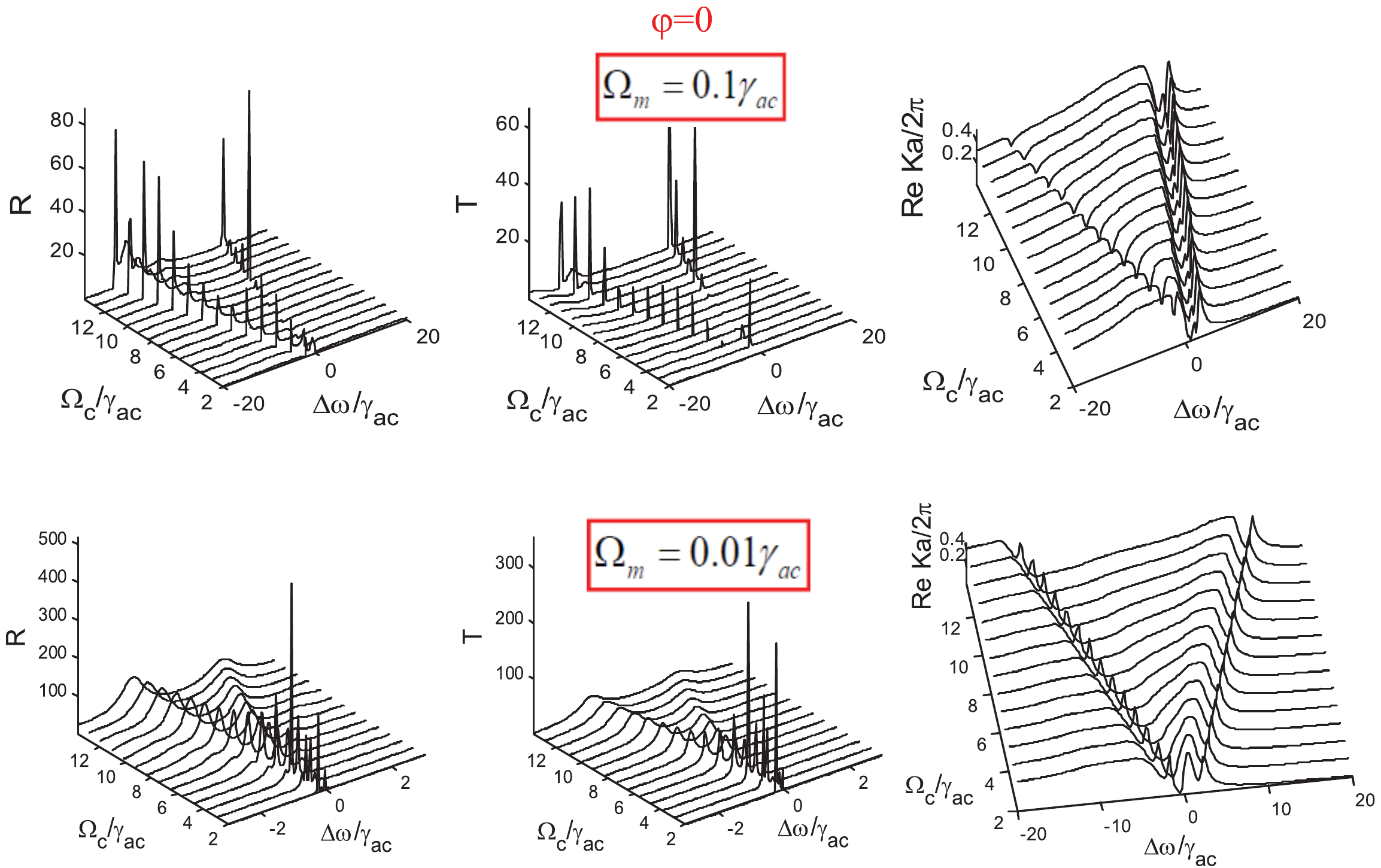}
\caption{Plot of the reflection, transmission and dispersion versus the Rabi frequency of the control field $\Omega_{c}$ for a) $\Omega _{m}=0.1\protect\gamma _{ac}$ and b) $\Omega _{m}=0.01\protect\gamma _{ac}$. We use $\Omega _{p}=0.1\protect\gamma _{ac}$, $\protect\rho _{cc}=0.0296$, $\protect\rho _{aa}=0.0155$ and other parameters are the same as in Fig. 2. The shift in the peaks becomes larger as $\Omega_m$ increases.}\label{Fig4}
\end{figure*}
The one-dimensional photonic crystal (superlattice) shown in Fig. \ref
{scheme} is composed of a finite number of dielectric-phaseonium pairs.
Multiple reflections and transmissions of a probe field E$_{0}$ through the
multilayers give the reflected field E$_{\text{r}}$=r\,E$_{0}$ and the transmitted
field E$_{\text{r}}$=t\,E$_{0}$ where the reflection and transmission coefficients are\cite
{Born}
\begin{equation}
r=\frac{(M_{11}+M_{12}p_{f})p_{i}-(M_{21}+M_{22}p_{f})}{(M_{11}+M_{12}p_{f})p_{i}+(M_{21}+M_{22}p_{f})},
\end{equation}
\begin{equation}
t=\frac{s_{if}2p_{i}}{(M_{11}+M_{12}p_{f})p_{i}+(M_{21}+M_{22}p_{f})},
\end{equation}
where $p_{j}=\zeta k_{jz}/\omega c$, $\zeta =\sqrt{\varepsilon _{0}/\mu _{0}}
$, $s_{if}$ $=1$ for TE-mode and $p_{j}=k_{jz}/\zeta k_{j}\sqrt{\varepsilon
_{j}}$, $\sqrt{\epsilon _{i}/\epsilon _{f}}$ for TM-mode ($\epsilon _{i}$
and $\epsilon _{f}$ are the dielectric constants at the initial and final
medium traversed by the probe.), $M_{ij}$ are the
components of the matrix
\begin{equation}
M=\left(
\begin{array}{cc}
m_{11}^{\prime }U_{N-1}-U_{N-2} & m_{12}^{\prime }U_{N-1} \\
m_{21}^{\prime }U_{N-1} & m_{22}^{\prime }U_{N-1}-U_{N-2}
\end{array}
\right)
\end{equation}
$m_{ij}^{\prime }$ are components of $2\times 2$ matrix $m^{\prime
}=(m_{2}m_{1})^{-1}$,
\begin{equation}
m_{j}=\left(
\begin{array}{cc}
\cos \beta _{j} & i\frac{\sin \beta _{j}}{p_{j}} \\
ip_{j}\sin \beta _{j} & \cos \beta _{j}
\end{array}
\right) ,j=1\,\,\text{or}\,\,2
\end{equation}
with 
\begin{equation}
\begin{split}
U_{N}=\frac{\sin (N+1)ka}{\sin ka},\\
\beta _{1}=k_{1z}d,\,\beta_{2}=k_{2z}s,\\
k_{jz}=\frac{\nu }{c}\sqrt{\epsilon _{j}-\epsilon _{i}\sin^{2}\theta }
\end{split}
\end{equation}
and $N$ is the number of phaseonium -dielectric layers. Also,
note that $M=(m^{\prime })^{N-1}$.

We note that the effective medium theory for susceptibility is only valid if
the number of periods is very large and the thickness of the two layers are
much smaller than the wavelength. For finite number of layers and the layers
thickness in the order of wavelength, the present theory is exact and the
effective medium theory is not a good approximation.

The results for the reflection, transmission for single interface, double
interface and superlattice are shown in Fig. \ref{Fig2}. We choose the
control laser to be resonant to focus on results with clearer features.
Figure  \ref{Fig2}b shows the usual two EIT peaks for single interface. Only
for the double interface (Fig. \ref{Fig2}c) and superlattice(Fig. \ref{Fig2}%
d), R and T can exceed unity since the probe field is amplified by acquiring
extra energy from the control field. From Figs. \ref{Fig2}b and c, for
single and double interface we see that the reflection peaks appear around
the two EIT resonances while the plateau of the transmission spectra lies
between the resonances. The two EIT peaks corresponding to absorption give
large reflection while the EIT window between the peaks gives high
transmission if it does not fall in the photonic bandgap of the
superlattice. However, there is a narrow transmission peak for double
interface which corresponds to the dispersion and absorption in Fig. \ref
{Fig2}a, is the result of double reflection at the two interfaces. In
general, multiple reflections lead to sharper reflection and transmission
peaks. However, for the superlattice structure (see Fig. \ref{Fig2}d) with
the input energy from the control laser, there is a broad peak of both R and
T are between the two EIT resonances. This is due to linear gain effect
which has covered up the fine peaks from multiple reflections.

To study the effect of both the amplitude and the phase of microwave field
we plotted (in Fig. \ref{Fig3}) the dispersion, reflection and transmission
as a function of Rabi frequency $|\Omega _{m}|$ for two choices of $\varphi
=0$ and $\pi /4$ where the distinct effect of the phase can be clearly seen
from the results shown in Fig. 3. For $\varphi =0$ the two (twin) EIT peaks
are shifted to lower frequency as $|\Omega _{m}|$ increases to around $%
\Delta \omega /\gamma _{ac}=0.1$. At this point, a small peak at higher
frequency appears but the location of this peak does not shift with $|\Omega
_{m}|$. Here, the EIT peaks becomes a single peak with the same size as the
peak at higher frequency. As $|\Omega _{m}|>$ $|\Omega _{c}|$ the separation
and width of the twin peaks increase with $|\Omega _{m}|$. Such peaks do not
appear in the absence of the microwave field. For $\varphi =\pi /4$, the
Rabi frequency becomes complex and we find the tall and narrow peaks in $R$
and $T$ around $|\Omega _{m}|\simeq 0.1\gamma _{ac}$ . It is important to
mention here that the fine resonant features exist only if $n_{ac}$ is
finite such that additional resonances are contributed by the term $\Gamma
_{cb}n_{ac}/\Gamma _{ac}$.

In Fig. \ref{Fig4}, we have plotted the dispersion, reflection and
transmission as a function of Rabi frequency $\Omega _{c}$ for three choices
of $\Omega _{m}=0.1\gamma _{ac}$ (upper) and $0.01\gamma _{ac}$ (lower). A
sufficiently large microwave field can promote/enhance the separation
between the twin EIT peaks in the transmission and reflection spectra as $\
\Omega _{c}$ increases. In particular the case $|\Omega _{m}|=0.1\gamma
_{ac}=|\Omega _{p}|$ transforms the lower frequency EIT peak into
interesting double peaks, one narrow and one broad when $\Omega _{c}$ \ is
sufficiently large.
\section{conclusion}
In summary, we have investigated the effects of microwave coupling along the
dipole forbidden (Raman) transition, on the quantum resonances in photonics
crystal with electromagnetic transparency. We envision new applications
based on our study. For example, the ultra-narrow transmission peak could be
useful in high-precision spectroscopy as well as ultraslow light buffer or
optical memory in all-photonic circuits. The narrow peak with high
transmission in the multilayer structure with finite gain could be used for
ultra narrow wavelength selection. 

Recently controlling Raman and sub-Raman
generation with microwave field has also been proposed\cite{Jha13}. The
presence of gain in active structures has been used to compensate for
absorption loss, promoting the practical use of quantum coherence in
metamaterials and photonic crystals to a wider domain. Such ideas have given
the birth of novel devices like nanolaser\cite{spaser1,Seidel05,Noginov09}
and recently coherence effects have also been reported\cite
{Dorfman13,PKJha13} in such configurations. Incorporating microwave along
with laser in such devices can bring additional degree of freedom for
enhancing the gain and controlling resonances. In particular NV center in diamond, coupled to surface plasmon (both localized and propagating) are good candidate for the realization of microwave induced control in plasmonics\cite{Huck11}. We note that extension of this work with microwave field beyond the system investigated here into the hot field of plasmonics would be beneficial. One major challenge for coherent control in plasmonics is to use a drive field which does not affect (heat up) the nanostructure. In general the plasmon resonances have bandwidth $\sim$100nm. Thus, microwave(maser) can be partially (if not completely) decoupled from the plasmon resonance and serve as an external control parameter. In general such approach also opens the door for nano-photonic devices to be merged with microwave telecommunication devices. However, for the experimental realization of the results presented here by the atomic physics/ quantum optics community, alkali metals like Rb,Na, K in a dielectric host are the feasible candidates. 
\section{Acknowledgement}
This work is supported by University of Malaya (UM)/Ministry of Higher
Education (MOHE) High Impact Research (HIR) programme Grant No.
A-000004-50001. P. K. Jha also acknowledges Herman F. Heep and Minnie Belle
Heep Texas A\&M University Endowed Fund held and administered by the Texas A\&M
Foundation and Robert A. Welch Foundation for financial support.


\begin{thebibliography}{99}
\bibitem{Tewari_PRL_86} S.P. Tewari and G.S. Agarwal, \prl {\bf 56}, 1811 (1986).
\bibitem{Harris90}S.E. Harris, J.E. Field, and A. Imam\u oglu, \prl {\bf 64}, 1107(1990)
 \bibitem{Boller91} K.-J. Boller, A. Imam\u oglu, and S.E. Harris, \prl {\bf 66}, 2593 (1991)
 \bibitem{Field91} J.E. Field, K.H. Hahn, and S.E. Harris, \prl {\bf 67}, 3062 (1991).
\bibitem{Schmidt99}H. Schmidt and A. Imamoglu, Opt. Lett. 21, 1936 (1996).
\bibitem{Harris99} S.E. Harris and L.V. Hau, Phys. Rev. Lett. 82, 4611 (1999).
\bibitem{Wang01} H. Wang, D. Goorskey, and M. Xiao, Phys. Rev. Lett. 87, 073601 (2001)
\bibitem{LWI1}O. Kocharovskaya and Ya. I. Khanin, JETP Lett. 48, 581(1988).
\bibitem{LWI2}S. E. Harris, Phys. Rev. Lett. 62, 1033(1989).
\bibitem{LWI3}M. O. Scully, S. Y. Zhu, and A. Gavrielides, Phys. Rev. Lett. 62, 2813(1989).
\bibitem{Jain96} M. Jain, H. Xia, G. Y. Yin, A. J. Merriam, and S. E. Harris, Phys. Rev. Lett. 77, 4326 (1996)
\bibitem{Jain93}M. Jain, J. E. Field, and G. Y. Yin, Opt. Lett. 18, 998 (1993).
\bibitem{Scully02}M. O. Scully, G. W. Kattawar, P. R. Lucht, T. Opatrny, H. Pilloff, A. Rebane, A. V. Sokolov, and M. S. Zubairy, Proc. Natl. Acad. Sci. U.S.A. 99, 10994 (2002).
\bibitem{Sari04}Z. E. Sariyanni and Y. Rostovtsev, J. Mod. Opt. 51, 2637 (2004).
\bibitem{Jha12}P. K. Jha, A. A. Svidzinsky and M. O. Scully, Laser Phys. Lett. 9, 368(2012).
\bibitem{Sete12}E. A. Sete, A. A. Svidzinsky, Y. V. Rostovtsev, H. Eleuch, P. K. Jha, S. Suckewer, and M. O. Scully, IEEE J. Sel. Top. Quantum Electron., \textbf{18}, 541 (2012). 
\bibitem{CRU}L. Yuan, A. A. Lanin, P. K. Jha, A. J. Traverso, D. V. Voronine, K. E. Dorfman, A. B. Fedotov, G. R. Welch, A. V. Sokolov, A. M. Zheltikov, and M. O. Scully, Laser Phys. Lett. 8, 736 (2011).
\bibitem{JhaAPL12} P. K. Jha,  K. E. Dorfman, Z. Yi, L. Yuan, Y. V. Rostovtsev, V. A. Sautenkov, G. R. Welch, A. M. Zheltikov, and M. O. Scully, Appl. Phys. Lett. 101, 091107 (2012).
  \bibitem{Dorfman13}K. E. Dorfman, P. K. Jha, D. V. Voronine, P. Genevet, F. Capasso and M. O. Scully (unpblished) arXiv:1212.523.
  \bibitem{PKJha13}P. K. Jha, X. Yin and X. Zhang, Appl. Phys. Lett. 102, 091111(2013).
\bibitem{Wilson05}E. A. Wilson, N. B. Manson, C. Wei, and L. Yang, Phys. Rev. A 72, 063813 (2005)
\bibitem{Kosa00} D. V. Kosachiov and E. A. Korsunsky, Eur. Phys. J. D 11, 457 (2000).
\bibitem{Joo10}J. Joo, J. Bourassa, A. Blais, and B. C. Sanders, Phys. Rev. Lett. 105, 073601 (2010).
\bibitem{Sahrai05}M. Sahrai, H. Tajalli, K. T. Kapale, and M. S. Zubairy, Phys. Rev. A 72, 013820 (2005).
\bibitem{Murali04} K. V. Murali, Z. Dutton, W. D. Oliver, D. S. Crankshaw, and T. P. Orlando Phys. Rev. Lett. 93 087003(2004).
\bibitem{Luo10}B. Luo, Y. Liu, and H. Guo, Opt. Lett. 35, 64(2010).
\bibitem{Jia10}W. Z. Jia and L. F. Wei, Phys. Rev. A 82, 013808(2010).
\bibitem{Novikova07}I. Novikova, A. V. Gorshkov, D. F. Phillips, A. S. S¿rensen, M. D. Lukin, and R. L. Walsworth, Phys. Rev. Lett. 98, 243602 (2007).  
\bibitem{Novikova07b}I. Novikova, David F. Phillips, and Ronald L. Walsworth, Phys. Rev. Lett. 99, 173604 (2007). 
\bibitem{Gorshkov07}A. V. Gorshkov, A. Andre, M. D. Lukin, and A. S. Sorensen, Phys. Rev. A 76, 033804 (2007); 76, 033805 (2007); 76, 033806 (2007).
\bibitem{Jha13}P. K. Jha, S. Das and T. N. Dey (unpublished) arXiv:1210.2356
\bibitem{Ooi00}  C. H. Raymond Ooi, T. C. Au Yeung, C. H. Kam, and T. K. Lim, Phys. Rev. B 61, 5920 (2000)
\bibitem{Ooi2010} C. H. Raymond Ooi and C. H. Kam, J. Opt.Soc. Am. B 27, 458 (2010).
\bibitem{Scully92}  M. O. Scully, Phys. Rep. 219, 191 (1992).
\bibitem{Ooi10}  C. H. Raymond Ooi, and C. H. Kam, Phys. Rev. B \textbf{81},195119 (2010)
\bibitem{And05} M. Bajcsy, A. S. Zibrov, and M. D. Lukin, Nature (London) 426, 638 (2003).
\bibitem{super T}  C.H.Raymond Ooi, and Q. Gong, J Appl. Phys. \textbf{110}, 063513(2011).
\bibitem{Li09}  H. Li , V. A. Sautenkov, Y. V. Rostovtsev, G. R. Welch, P. R. Hemmer, and M. O. Scully,  Phys. Rev. A \textbf{80}, 023820 (2009).
\bibitem{Yariv}A. Yariv and P. Yeh, \textit{Optical Waves in Crystals.} J. Wiley, New York, 1984.
\bibitem{Thesis12} P. K. Jha, Ph.D. Dissertation (Texas A$\&$M University, College Station, 2012).
\bibitem{Jha11a}  P. K. Jha, Y. V. Rostovtsev, H. Li, V. A. Sautenkov and M.O.Scully, Phys. Rev. A 83, 033404 (2011)
\bibitem{Jha11b} P. K. Jha, H. Li, V. A. Sautenkov, Y.V.Rostovtsev and M.O.Scully, Opt. Commun. 284, 2538 (2011).
\bibitem{Born}  M. Born and E. Wolf, Principles of Optics, 7th ed.
(Cambridge University Press, United Kingdom, 1999).
\bibitem{spaser1}D. J. Bergman, and M. I. Stockman, Phys. Rev. Lett. \textbf{90}, 027402 (2003).
     \bibitem{Seidel05}J. Seidel, S. Grafstroem, and L. Eng, Phys. Rev. Lett. \textbf{94}, 177401 (2005).
\bibitem{Noginov09} M. A. Noginov, G. Zhu, A. M. Belgrave, R. Bakker, V. M. Shalaev, E. E. Narimanov, S. Stout, E. Herz, T. Suteewong and U. Wiesner, Nature \textbf{460}, 1110(2009).
\bibitem{Huck11}A. Huck, S. Kumar, A. Shakoor, and U. L. Andersen, Phys. Rev. Lett. \textbf{106}, 096801 (2011)
\end{thebibliography}
\end{document}